# India's rank and global share in scientific research: *how data sourced from different databases can produce varying outcomes?*

**Prashasti Singh[1], Vivek Kumar Singh[*1], Parveen Arora[2] & Sujit Bhattacharya[3]**

[1]Department of Computer Science, Banaras Hindu University, Varanasi-221 005, India.
[2]Department of Science and Technology, Govt of India, New Delhi-110 016, India.
[3]CSIR- National Institute of Science Technology and Development Studies, New Delhi-110 012, India.

**Abstract:** India is emerging as a major knowledge producer of the world in terms of proportionate share of global research output and the overall research productivity rank. Many recent reports, both of commissioned studies from Government of India as well as independent international agencies, show India at different ranks of global research productivity (variations as large as from 3rd to 9th place). The paper examines this contradiction; tries to analyse as to why different reports places India at different ranks and what may be the reasons thereof. The research output data for India, along with the ten most productive countries in the world, is analysed from three major scholarly databases- Web of Science, Scopus and Dimensions for this purpose. Results show that both, the endogenous factors (such as database coverage variation and different subject classification schemes) and the exogenous factors (such as subject selection and publication counting methodology) cause the variations in different reports. This paper reports first part of the analysis, focusing mainly on variations due to use of data from different databases. The policy implications of the study are also discussed.

## Introduction

During the last two decades, India has not only emerged as a major player in geo-political and economic landscape of the world, but has also significantly improved its place among major knowledge producers. India is growing, both in terms of absolute research output as well as in its proportionate share to global research output. Various reports from multiple agencies have highlighted this fact. In fact, the 2020 report on Science and Engineering Indicators by National Science Foundation (NSF)[1], have ranked India as the 3rd biggest knowledge producer, in the area of Science and Engineering. Similarly, the most recent report on Research and Development Statistics[2], based on the studies commissioned by Department of Science and Technology to Clarivate Analytics (owner of Web of Science database) and Elsevier (owner of Scopus database), shows that India ranks at 9th (Web of Science data) and 5th place (Scopus database), for the year 2018, in terms of global research output. Thus, if we look at these three findings, we find India ranked at 3rd, 5th and 9th places on research output, in the year 2018.

Though India's emergence as a major knowledge producer is being acknowledged in various studies and reports, however, at the same time the varying evidence in different reports also create some confusion about India's actual rank on global research output. It is in this context that this article tries to analyse as to why different studies/ reports may be showing India at different ranks of global research output. It is observed that both, the endogenous factors (such as database coverage variation and different subject classification schemes) and the exogenous factors (such as subject selection and publication counting methodology) cause the variations in different reports. This paper reports findings of the first part of the analysis, focusing mainly on variations due to use of data from different databases. The research output data (for India and the 10 most productive countries) is obtained from the three popularly used scholarly databases- Web of Science, Scopus and Dimensions for the period 2010-19, and analysed. Results obtained show that use of data from different databases produce significantly different outcomes.

## Objectives

This article attempts to analyse the endogenous factors responsible for variations observed in India's global research output, rank and proportionate share. While examining the variations observed in India's ranking and exploring the reasons therein, the paper also examines the publication patterns of ten most productive countries to see whether this variation is more specific to India or is seen for other countries as well.

**Data & Method**

The research output data for India and the ten most productive countries of the world is obtained from the three main scholarly databases- Web of Science, Scopus and Dimensions, for the period 2010 to 2019. To get the data from Web of Science, a search query PY= (2010-2019) was used to get the publication counts, which was then limited to selected countries. In Scopus the query was set to PUBYEAR > 2009 AND PUBYEAR < 2020 for obtaining the corresponding data. Similarly, in Dimensions databases, the corresponding data was obtained by using API queries with publication year filters set to the range 2010-2019. We obtained full publication data, comprising of all document types, for the countries due to following two reasons: (a) different databases categorize publications in different document types, with no direct correspondence between them, and (b) several standard reports, such as the Compendium of Bibliometric Science Indicators report by OECD[3] suggest to use full data for document types for scientometric assessment exercise involving countries. We have taken a broader definition of scientific research to include research output in 'Arts & Humanities' and 'Social Science' disciplines, as suggested in the OECD Compendium. However, it may also be noted that 'Arts & Humanities' and 'Social Science' disciplines constitute a small amount of publication volume as compared to Science, Technology, Engineering and Medicine fields.

In order to have a more detailed analysis of the variations in India's output in different databases, detailed metadata for publications from India for the year 2016 (as an example year, and perhaps the most recent year with stable data in the databases) was also downloaded from all the three databases. For the publication year 2016, Web of Science had 76,836 publication records indexed, Scopus had 154,858 records indexed, and Dimensions had 136,089 records indexed. These counts include all document types and subject areas. The data was processed and filtered by removing duplicates and erroneous records, which left us with 67,367 unique records from Web of Science, 96,908 unique records from Scopus, and 123,738 unique records from Dimensions. Out of these, we used publication records for document types- 'article' and 'review' to identify the unique and overlapping set of journals from which the year 2016 publications are drawn by the databases.

The method for analysis comprised mainly of quantitative and computational approaches. Programs were written in Python programming language to pre-process the data and to compute different values and measures. The results are shown in tables and figures, drawn mainly by using Python and Excel functions and utilities.

*First of all*, India's total research output, rank and proportionate share was computed for all the years during the period 2010-19. Thereafter, the annual growth rate and compounded annual growth rate (CAGR) for the whole period was computed for data from all the three databases.

*Secondly*, the detailed publication records for India, from all the three databases for the publication year 2016, were analysed to find out overlapping and unique records in the three databases. This was done for document types 'article' and 'review' in different databases, mainly for the purpose of identifying unique and overlapping journal indexing of the three databases for the year 2016.

*Thirdly*, research output numbers of India and the ten most productive countries obtained from the three databases, are analysed. The relative research output ranks and global share of different countries are computed to see if across-database variations are observed in case of other countries.

*Finally*, the subject-wise distribution of Indian research output for the year 2016 is analysed to understand how different databases have altogether different subject classification schemes, and how it makes comparisons across the databases more difficult.

**Results**

The analytical results are organized into four major parts. First of all, India's research output, global rank and share is presented. Secondly, a detailed analysis of publication records for India for the year 2016 is shown to see how and why the variations in India's productivity levels in different databases may be there. Thirdly, variations in productivity levels of ten most productive countries as per data from different databases is presented. Finally, the subject area distribution of India's research output in all the three databases is shown along with its implications.

**India's research output, rank and global share**

We first look at research output for India for ten-year period (2010-19) from all the three databases. Table 1 presents the number of publications, annual growth rate, CAGR and India's rank in global research output, computed for data from all the three databases. It is observed that India's research output has grown during 2010 to 2019, with CAGR values of 6.7% (Web of Science data), 8.1% (Scopus data), and 10% (Dimensions data). It is also interesting to observe that, as per data from Web of Science, India has improved its global rank from being 12th place in 2010 to 10th place since 2017 onwards. The data from Scopus places India at a higher rank, improving from 9th place in 2010 to 5th place since 2014 onwards. The Dimensions data shows India improving from 9th place in 2010 to 5th place in 2019. Thus, as per the three databases, India's global ranks in research productivity in 2019 are 10 (Web of Science) and 5 (both Scopus and Dimensions). It is equally interesting to observe that the volume of research output for India indexed in the three databases vary significantly (as large as 104% between Web of Science and Scopus in the year 2018 and 81% between Web of Science and Dimensions). In general, Scopus and Dimensions are closer in values as compared to Web of Science.

We have also computed India's proportionate share to global research output during 2010-2019 period, as per data from all the three databases. Fig. 1 shows the trend in India's proportionate share to global research output, in all the three databases. It is observed that, as per Scopus data, India's global share of research output has increased from 3.3% (in 2010) to 5.65% (in 2019). As per Web of Science data, the increase is from 2.71% (in 2010) to 3.86% (in 2019). The values for Dimensions database show an increase from 2.15% (in 2010) to 3.26% (in 2019). Thus, it is seen that use of different databases not only present different evidence for research output volume and global rank, but also present different values of global research output share. As in 2019, India's global research output share stands between 3.26% to 5.65%, as per data from different databases. However, leaving aside the variations, the common trend in data across all the three databases is that India's global research output share is on the rise.

**Table 1: India's output, annual growth rate & global rank (2010-2019)**

| Year | Web of Science | | | Scopus | | | Dimensions | | |
|---|---|---|---|---|---|---|---|---|---|
| | Output | AGR | Rank | Output | AGR | Rank | Output | AGR | Rank |
| 2010 | 48,386 | - | 12 | 81,395 | - | 9 | 65,948 | - | 9 |
| 2011 | 52,724 | 8.97 | 11 | 98,835 | 21.43 | 7 | 69,311 | 5.1 | 9 |
| 2012 | 56,082 | 6.37 | 12 | 109,479 | 10.77 | 7 | 82,067 | 18.4 | 9 |
| 2013 | 62,383 | 11.24 | 11 | 117,640 | 7.45 | 7 | 98,588 | 20.13 | 7 |
| 2014 | 68,071 | 9.12 | 11 | 133,276 | 13.29 | 5 | 117,052 | 18.73 | 7 |
| 2015 | 70,879 | 4.13 | 11 | 141,912 | 6.48 | 5 | 124,151 | 6.06 | 6 |
| 2016 | 76,836 | 8.4 | 11 | 154,858 | 9.12 | 5 | 136,089 | 9.62 | 6 |
| 2017 | 79,818 | 3.88 | 10 | 156,371 | 0.98 | 5 | 148,916 | 9.43 | 6 |
| 2018 | 85,212 | 6.76 | 10 | 174,629 | 11.68 | 5 | 154,716 | 3.89 | 6 |
| 2019 | 92,435 | 8.48 | 10 | 176,925 | 1.31 | 5 | 171,974 | 11.15 | 5 |
| CAGR | 6.7% | | | 8.1% | | | 10% | | |

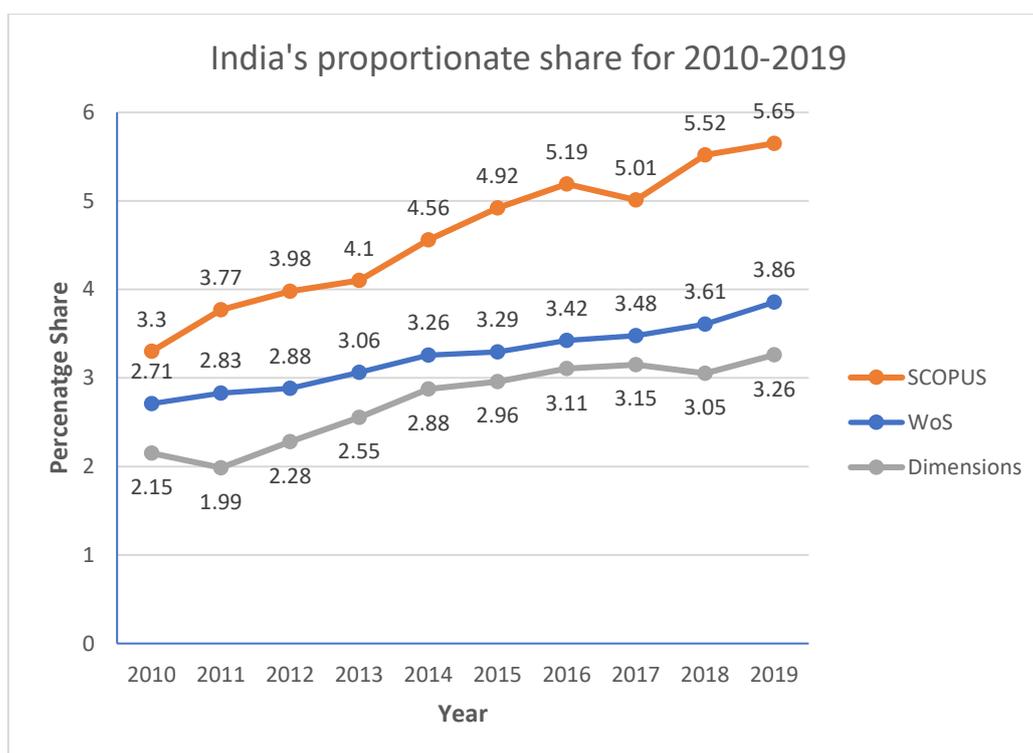

**Fig. 1- India's proportionate share in global research output as seen in three databases**

**Detailed analysis of India's research output for 2016**

To understand as to why publication volumes for India vary across the databases, a detailed analysis of research output data for India for the publication year 2016 was done. *First of all*, it is important to examine the data pertaining to distribution of document types in the three databases. Table 2 shows the document type distribution of publications records for the year 2016 in all the three databases. It can be observed that Web of Science database has only 1,025 conference papers indexed, as against 20,189 in Scopus and 21,182 in Dimensions. Therefore, one major reason for variation in publication volume is the varying amount of coverage of conference papers in Web of Science and the other two databases. *Secondly*, it is also observed that number of publication records of the type 'article' also vary significantly in the three databases, ranging from 57,844 in Web of Science to 66,955 in Scopus and 94,387 in Dimensions.

Given that the three databases not only differ in their conference coverage but also in number of journal articles indexed in them, it became necessary to look at the difference in volume of journal articles indexed by them. For this purpose, we selected two main document types- 'article' and 'review', both of which have a journal as publication source. Fig. 2 shows the overlapping and unique journal publication records indexed in the three databases. It is observed that Web of Science has a total of 61,089 publication records (57,844 articles and 3,245 reviews). Scopus database has a total of 70,599 publication records (66,955 articles and 3,644 reviews). Dimensions database has 94,387 articles (Dimensions does not have document type 'review' and apparently indexes that as document type 'article'). Out of the article and review document types considered, 40,135 publication records are common in all the three databases. Excluding the publication records common in all three databases, it is found that 10,510 publication records are common in Web of Science and Scopus; 8,790 publication records are common between Scopus and Dimensions; and 7,358 publication records are common between Dimensions and Web of Science. Thus, 82.9% of publication records indexed in Web of Science are also indexed in Scopus, and 77.7% are also indexed in Dimensions. In case of Scopus database, 71.7% of its publication records are indexed in Web of Science and 69.2% publication records are indexed in Dimensions. In case of Dimensions database, 50.3% of its publication records are found indexed in Web of Science and 51.8% also are indexed in Scopus. The Web of Science database has

3,086 uniquely indexed publication records, Scopus has 11,164 uniquely indexed publication records, and Dimensions has 38,104 uniquely indexed publication records. Thus, it is seen that not only the three databases vary in coverage of different types of document types (such as conference proceedings), but also in publication records of the same type (as seen with document types 'article' and 'reviews', both of which have journal as a publication source). This implies that the three databases have significantly different coverage of journals.

In order to understand variation in journal coverage of the three databases, we extracted all distinct journal names for the publication records for the year 2016. Fig. 3 shows the overlapping and unique journals covered in the three databases for the Indian publication records (document types 'article' and 'review') for the year 2016. It is observed that the 61,089 publication records (article + review) in Web of Science are published in 6,116 distinct journals. Similarly, the 70,599 publication records (article + review) in Scopus are drawn from 7,776 distinct journals covered by Scopus. The 94,387 articles in Dimensions are drawn from 8,702 distinct journals. Thus, the article coverage in Scopus is coming from additional 1,660 journals as compared to Web of Science, and coverage in Dimensions is coming from additional 2,586 journals as compared to Web of Science. A total of 3,697 journals are found common in all the three databases. There are also many journals that are uniquely covered in just one database. For example, 955 journals (from which the 2016 research output is drawn) are only covered in Web of Science; 1,896 journals are uniquely covered in Scopus; and 2,890 journals are uniquely covered in Dimensions. Thus, given that the different databases have different number of journals covered by them, selecting publications records for the same document type (say 'article') will also produce different evidence of research outputs.

Table 2: Document type distribution in the three databases for India (2016)

| Document Type | Web of Science | Scopus | Dimensions |
|---|---|---|---|
| Article | 57,844 | 66,955 | 94,387 |
| Conference Proceedings Paper (Conferences) | 1,025 | 20,189 | 21,182 |
| Biographical Item | 22 | — | — |
| Book | — | 187 | — |
| Book Chapter | 19 | 2,748 | 8,014 |
| Book Review | 214 | — | — |
| Correction | 314 | — | — |
| Editorial | 1,552 | 589 | — |
| Erratum | — | 248 | — |
| Letter | 1,590 | 1,602 | — |
| Meeting Abstract | 1,458 | 1 | — |
| Note | — | 631 | — |
| News Item | 37 | — | — |
| Preprint | — | — | 155 |
| Reprint | 1 | — | — |
| Retraction | 44 | 21 | — |
| Review | 3,245 | 3,644 | — |
| Short Survey | — | 93 | — |
| Software Review | 2 | — | — |
| TOTAL | 67,367 | 96,908 | 123,738 |

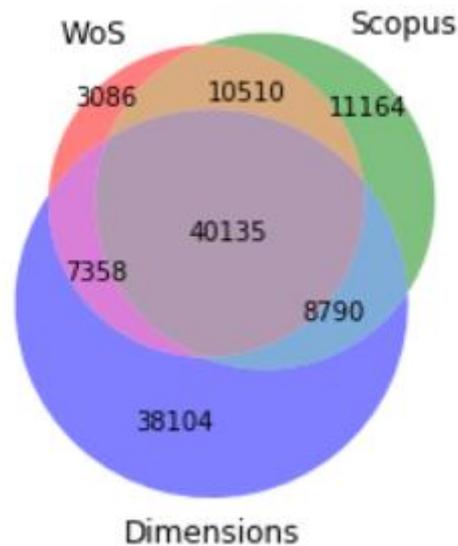

**Fig.2- Overlapping & unique articles for India for 2016 in the three databases**

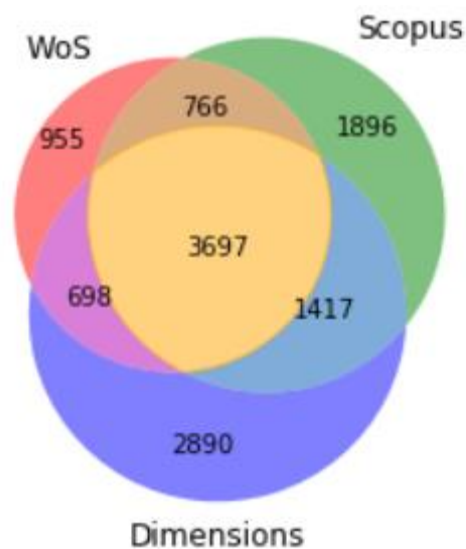

**Fig. 3- Overlapping and unique set of journals in which Indian output for 2016 appears**

**Variations in research productivity levels of different countries**

Since we observed variations in research output volume and rank of India in data drawn from the three different databases; we tried to analyse whether such variations are also seen for other countries. The overall research output data for 10 most productive countries and their relative research output levels in the three databases were analysed. Table 3 shows the overall research output, global rank and share of the 10 most productive countries, and India. It is observed that USA, China, UK and Germany are the four countries ranked higher, in the same order of research output, in all the three databases. However, the number of records for all these four countries vary significantly in the three databases. For example, China has 2,994,155 publication records indexed in Web of Science, 4,865,151 publication records indexed in Scopus, and 3,799,383 publication records indexed in Dimensions, for the period 2010-19. The variations in number of records for China in the three databases is thus close to 1 million.

It is further observed that Web of Science data shows Japan, France, Canada, Italy, Spain, Australia and India at research output ranks from 5 to 11, respectively. However, according to Scopus, India stands above Japan, France, Canada, Italy, Spain and Australia. Further, relative ranks of research output in Scopus differs from Web of Science, with Canada, Italy, Spain and Australia having different relative ranks in the two databases. Similar variations are observed for data from Dimensions database as well. Here, Japan and France are above India, unlike being below India in Scopus. The relative rank of Canada and Italy are different in Scopus and Web of Science. When we look at proportionate global share of all these countries in different databases; it is seen that USA alone has significant variation, with 28.01% global share in Web of Science data, 23.07% global share in Scopus data, and 15.84% global share in Dimensions data. China varies with 14.14% global share in Web of Science, 16.8% in Scopus and 9.1% in Dimensions.

Table 3: Research output for highly productive countries (2010-2019)

| Rank | Web of Science | | | Scopus | | | Dimensions | | |
|---|---|---|---|---|---|---|---|---|---|
| | Country | Output | Global Share[a] | Country | Output | Global Share[b] | Country | Output | Global Share[c] |
| 1 | USA | 5,930,830 | 28.01% | USA | 6,671,185 | 23.07% | USA | 6,610,125 | 15.84% |
| 2 | China | 2,994,155 | 14.14% | China | 4,865,151 | 16.82% | China | 3,799,383 | 9.10% |
| 3 | UK | 1,475,674 | 6.97% | UK | 2,010,318 | 6.95% | UK | 1,967,828 | 4.71% |
| 4 | Germany | 1,366,593 | 6.45% | Germany | 1,735,983 | 6.00% | Germany | 1,803,772 | 4.32% |
| 5 | Japan | 997,622 | 4.71% | India | 1,345,320 | 4.65% | Japan | 1,513,116 | 3.63% |
| 6 | France | 920,247 | 4.35% | Japan | 1,324,302 | 4.58% | France | 1,181,208 | 2.83% |
| 7 | Canada | 892,093 | 4.21% | France | 1,193,743 | 4.13% | India | 1,168,812 | 2.80% |
| 8 | Italy | 860,330 | 4.06% | Italy | 1,084,908 | 3.75% | Canada | 1,036,389 | 2.48% |
| 9 | Spain | 736,849 | 3.48% | Canada | 1,050,984 | 3.63% | Italy | 1,016,990 | 2.44% |
| 10 | Australia | 769,428 | 3.63% | Australia | 931,857 | 3.22% | Australia | 888,698 | 2.13% |
| 11 | India | 692,826 | 3.27% | Spain | 893,644 | 3.09% | Spain | 868,771 | 2.08% |

[a] Corresponding to Total Research Output of the World in Web of Science (2010-19) = 21,172,341
[b] Corresponding to Total Research Output of the World in Scopus (2010-19) = 28,919,294
[c] Corresponding to Total Research Output of the World in Dimensions (2010-19) = 41,739,651

The variations in volume of research output, global share and rank in data from different databases is thus also observed for the most productive countries. It may be noted that full publication records (all document types) for all the countries are taken for comparison. This is done due to the fact that there is no clear and direct correspondence between document types in different databases. For example, Dimensions database does not have type 'review' and apparently indexes 'reviews' under document type 'article'. The holistic comparison may have limitations, mainly in terms of limited conference coverage and separate index for Books in Web of Science. However, the other two databases are more similarly organized in terms of their document type coverage, with same index covering all kinds of data. Nevertheless, the variations in publication volumes and global share of different countries are quite noticeable. Taking into account the fact that the number of publication records for the same document type also vary significantly (as shown in case of Indian data for 2016), it would not be unreasonable to expect that limiting the comparison to the publication records of the same document type in the three databases will still result in significant variations in research output, global share and rank of different countries.

**Subject area distribution of research output in different databases**

The third aspect that has been observed from analysing the publication records of the three databases is that each of them have a different subject classification scheme and classification granularity. Each of these databases follow different levels and types of subject classification of publication records. In order to understand this more clearly, the Indian research output for the year 2016 from all the three databases is observed with the corresponding disciplinary distribution. Figs 4-6 show the disciplinary distribution of publication records for India for the year 2016 as provided by Web of Science, Scopus and Dimensions, respectively.

It can be observed that according to Web of Science subject area distribution, the largest proportion of research output is in Chemistry (with all its sub-categories) with 11.69% share, followed by Engineering with 8.42% share. These are followed by Physics with 7.35% share and Material Science with 6.84% share. On the other hand, if we look at subject area distribution in Scopus data, the major shares are of Engineering with 13.82% share, Computer Science with 11.49% share and Medicine with 9.74% share. In case of Dimensions, the largest share is in Medical and Health Science with 26.29% share followed by Engineering with 17.65% share. Other major shares are of Chemical Science (11.94%) and Information and Computing Science (11.75%).

It may be noted that the subject areas in the three databases differ significantly in their composition of specific sub-categories. Therefore, drawing publication data from different databases is bound to produce different evidence of India's research strength in different subject areas. For example, it is seen that as per Web of Science data, out of total research output from India, 11.69% is in Chemistry; whereas proportionate share of Chemistry as per Scopus data is 6.32%, and as per Dimensions data is 11.94%. Similarly, for the research output in Computer Science, Web of Science data shows 2.11% share; whereas Scopus shows 11.94% share and Dimensions shows 11.75% share. All these databases thus produce significantly different evidence when it comes to research output in specific subject areas. This implies that if a research productivity assessment is to be done in a specific subject area, the most suitable database with best coverage may be selected. For example, for subject area Computer Science, using the Web of Science data may not be suitable due to its low coverage of the subject area, instead using Scopus and Dimensions may produce more informed assessments, either at the level of institutions or countries.

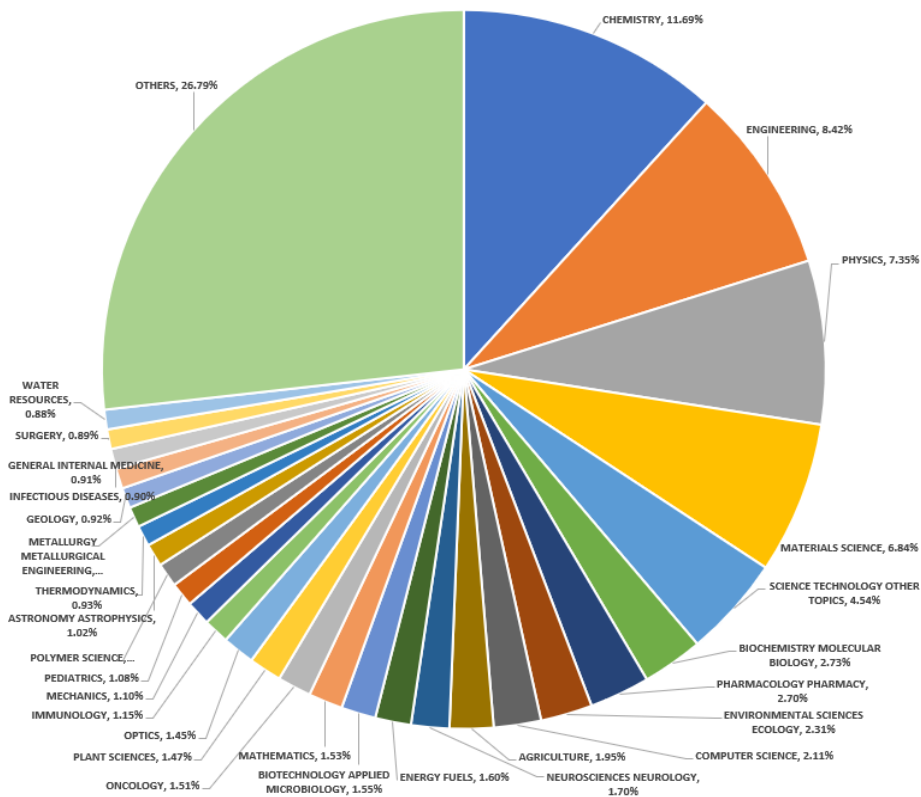

**Fig. 4- Disciplinary distribution of Indian research output as in Web of Science database (2016)**

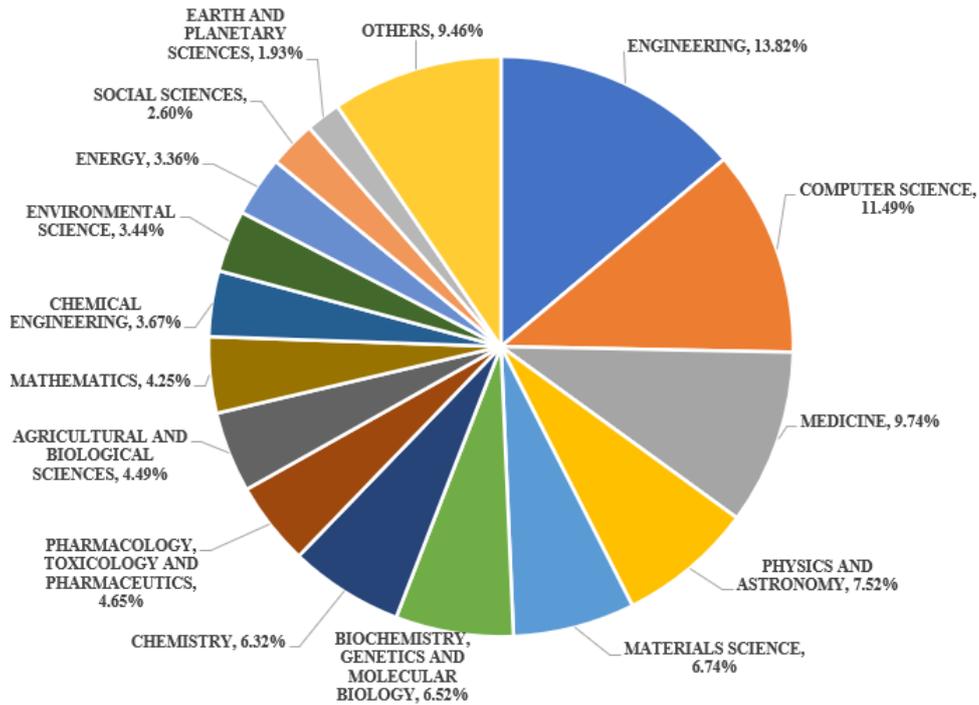

**Fig. 5- Disciplinary distribution of Indian research output as in Scopus database (2016)**

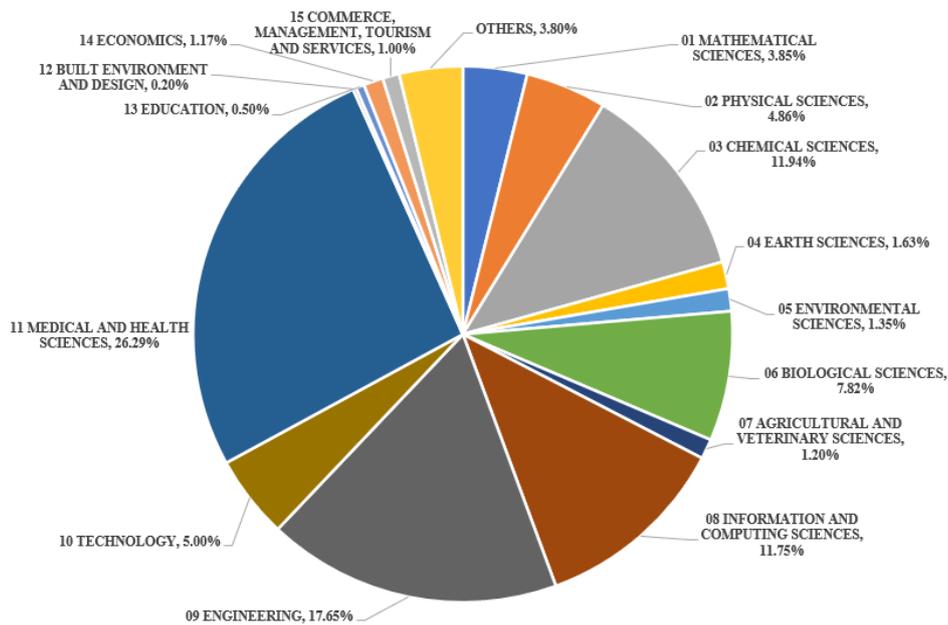

**Fig. 6- Disciplinary distribution of Indian research output as in Dimensions database (2016)**

The different subject classification schemes of different databases may cause further variations in research output volume, global share and rank of countries. To illustrate this, we analysed publication records for the ten most productive countries and India, for two subject areas, Computer Science and Agricultural & Veterinary Sciences. Table 4 shows the research output rank and the number of publication records for the subject Computer Science for the period 2010-19. It can be observed that China and USA are the two top ranked countries in order of research output in all the three databases.

However, below this level, the relative positions of countries vary significantly. For example, India is placed at rank 8 in Web of Science, but at rank 3 in Scopus and Dimensions both. Japan is not listed in top 11 most productive countries in Web of Science, whereas it is placed at rank 6 in Scopus and Dimensions both. The relative ordering of many country pairs also varies across databases. For example, in Web of Science UK is listed above Germany, whereas in Scopus and Dimensions it is listed below Germany. Table 5 shows the research output rank and the number of publication records for the subject Agricultural & Veterinary Science. It is observed that USA and China are placed at top two places, in order, in all the databases. However, countries like UK, Brazil etc. are placed in different orders in data of different databases. In this case too, India is placed at 6$^{th}$ place of research productivity in Web of Science and Scopus, but it is placed at 11$^{th}$ place in Dimensions.

The subject-specific data shows two interesting observations. First, that number of records for the same country for a specific subject area is found to be significantly different in data from the three databases. Secondly, the relative ranks of countries in the data for two different subjects is significantly different. For example, Brazil is not listed in Table 4 (Computer Science) but appears at prominent place in Table 5 (Agriculture & Veterinary Science). These observations thus produce following useful implications: (a) selection of different subsets of subjects for research assessment exercises will produce different results, and (b) selection of a suitable database is quite important for informed outcomes in subject-specific research assessment exercise.

**Table 4: Most productive countries in Computer Science research (2010-2019)**

| Rank | Web of Science | | Scopus | | Dimensions | |
| --- | --- | --- | --- | --- | --- | --- |
| | Country | Output | Country | Output | Country | Output |
| 1 | China | 123,584 | China | 799,267 | China | 447,151 |
| 2 | USA | 1,061,81 | USA | 643,654 | USA | 384,271 |
| 3 | UK | 29,347 | India | 251,049 | India | 137,416 |
| 4 | Germany | 24,767 | Germany | 213,710 | Germany | 124,340 |
| 5 | Spain | 24,344 | UK | 189,554 | UK | 113,859 |
| 6 | France | 24,325 | Japan | 169,350 | Japan | 97,832 |
| 7 | Canada | 22,094 | France | 156,791 | France | 88,897 |
| 8 | India | 20,859 | Italy | 130,790 | Italy | 75,474 |
| 9 | South Korea | 20,197 | Canada | 115,909 | Canada | 70,446 |
| 10 | Italy | 19,680 | South Korea | 108,480 | Spain | 62,631 |
| 11 | Taiwan | 18,102 | Spain | 106,952 | South Korea | 58,251 |

**Table 5: Most productive countries in Agricultural & Veterinary Sciences research (2010-2019)**

| Rank | Web of Science | | Scopus | | Dimensions | |
| --- | --- | --- | --- | --- | --- | --- |
| | Country | Output | Country | Output | Country | Output |
| 1 | USA | 202,948 | USA | 528,215 | USA | 98,696 |
| 2 | China | 148,857 | China | 337,536 | China | 46,890 |
| 3 | Brazil | 79,416 | UK | 151,719 | UK | 27,771 |
| 4 | Germany | 54,140 | Brazil | 146,918 | Brazil | 26,254 |
| 5 | Spain | 49,662 | Germany | 137,507 | Canada | 21,820 |
| 6 | India | 48,927 | India | 120,148 | Australia | 20,278 |
| 7 | Japan | 48,636 | Australia | 105,291 | Japan | 18,951 |
| 8 | Italy | 40,994 | Canada | 101,589 | Spain | 18,785 |
| 9 | Canada | 40,223 | France | 101,049 | Germany | 18,677 |
| 10 | Australia | 38,096 | Spain | 99,489 | Italy | 16,642 |
| 11 | UK | 37,991 | Japan | 96,880 | India | 15,960 |

**Discussion**

The analytical results show significant variations in volume and global share of research output of India as per data from different database. Interestingly similar variations are observed for other highly productive countries, if data is drawn from different databases. Not only the research output volumes vary, but the relative ranks of different countries vary significantly in data from different scholarly databases. This section discusses how the endogenous factors (database used and their varying subject classification schemes) cause these variations.

The analysis of publication records from the three databases show that they vary significantly in coverage of different types of publication records. For example, it is observed that Web of Science has poor coverage of conference proceedings as compared to Scopus and Dimensions. Therefore, any research assessment exercise that includes conference proceedings will produce significantly different results with change in database from Web of Science to Scopus or Dimensions. These coverage variations may thus be an important factor in variations in research output evidence from different databases.

In addition to variations in coverage of publication records of different types, it is also seen from the analytical results that different databases have varying coverage of publication records of the same type. For example, Web of Science covers journal articles from much lesser number of journals as compared to Scopus and Dimensions. In case of Indian research output ('article' and 'review' document types) for the year 2016, it is observed that Web of Science draws the publication records from 6,116 distinct journals, Scopus draws from 7,776 distinct journals and Dimensions draws from 8,702 distinct journals. Thus, the number of journals that are covered for 2016 research output for India in Scopus is about 27% more as compared to Web of Science. Similarly, in case of research output in Dimensions, the journals covered is about 42% more than Web of Science database. Variation in coverage of journals is also seen among Scopus and Dimensions. Therefore, it can be understood that use of data from different databases is bound to produce different evidence of research performance assessment, since they vary significantly not only in their coverage of different publication types (mainly journal and conferences) but also in coverage of same types of publications. A somewhat similar observation was also recorded in study on journal articles and review data by Mongeon & Paul-Hus (2016)[(4)], wherein they show that productivity ranks of 15 selected countries differ with change of database used from Web of Science to Scopus. Another recent study (Huang et al., 2020)[5] compared different bibliographic data sources with respect to robustness of University rankings, and observed coverage variations as a significant reason for varied outcomes.

Another important factor that may cause significant variations in findings of assessment exercises based on data drawn from different databases is the varying subject classification schemes of different databases. Different databases not only have different number of subject areas in which they classify the publication records, but they also adopt different approaches of assigning a publication record to a subject area. There are two different kinds of subject classification schemes used in these databases. Web of Science and Scopus use a source-based classification, in which each journal is permanently assigned to one or more subject categories. Thus, in both these databases, an article is assigned to a subject area which is permanently tagged with the journal in which it is published. However, even though both these databases follow source-based subject classification, they have different levels of subject classification granularity. The Web of Science has a 2-level subject classification scheme with 254 subject areas in which journals are classified (and hence the articles in these areas). Scopus has a 3-level subject classification scheme with 4 broad research areas, 27 subject categories, that are further divided into 334 minor subject categories. The Dimensions database uses an article-level subject classification, with 2-level subject classification scheme, comprising of 22 divisions and 157 groups in the divisions.

Taking into account the large-scale differences in number of subject categories and sub-categories used in different databases, it is difficult to have a direct correspondence between 254 subject areas of Web of Science with 334 minor subject categories in Scopus or 157 divisions of Dimensions. This implies that research assessment exercises for a specific subject area or a subset of subject areas will produce different outcomes if a different database is used. Further, since each database has distinct coverage of different subject areas, it may be advisable to select the most suitable database for subject-specific research assessment exercises.

The variations in outcomes of different studies can also be caused by some exogenous factors, even if they use the same database. One such example is the variation observed in report of commissioned study done by Elsevier for India and the NSF Science and Engineering Indicators report of 2020, both of which use data from Scopus. These variations usually occur due to use of data for selected disciplines and/ or due to use of different publication counting methods. The second part of this study (Singh *et al.*, 2020)[6] analyses in detail the impact of exogenous factors, mainly the impact of subject selection and publication counting method (whole or fractional counting) used.

**Conclusion**

The paper analyses the effect of varied coverage of scholarly databases on research output volume, global share and rank of India. Research output data is obtained from three popular scholarly databases (Web of Science, Scopus and Dimensions) and analysed. The analysis produces interesting observations, which lead us to useful and interesting conclusions.

*First*, the variations in research assessment reports are bound to happen if data is taken from different scholarly databases. The main reasons for this are varied coverage of different databases, both the publication records for different document types (such as different numbers of articles and conference papers indexed in each database) as well as publication records of the same document type (such as different number of journal articles covered in the three databases due to their varied journal coverage). *Secondly*, assessment exercises at country level should preferably be based on holistic publication data, rather than a subset of the data, for more informed and robust outcomes. *Thirdly*, the overall trend of growth in volume and global share of research output should be taken as the more important outcome of research assessment exercises, and over-emphasis on absolute ranks should always be avoided. *Finally*, given that different databases use different subject classification schemes, with no direct correspondence, and hence results of subject specific assessment exercises (involving a single or a subset of subjects) from different databases cannot be compared. It also implies that it is important to carefully select the most appropriate database for research assessment exercises in specific subjects.

This article mainly analysed the impact of endogenous factors on the variations in research assessment outcomes of different studies/ reports based on different databases, with impact of exogenous factors dealt with in second part of the study (Singh *et al.*, 2020)[6]. In this study, a holistic comparison involving all document types and subject areas is carried out, however, it may be interesting to further analyse the variations at different granularities involving selected document types and/ or subject areas. Similarly, the relative coverage of Indian journals in the three databases and their impact on outcomes of reports based on those databases, would also be an equally interesting exercise to pursue.